\documentclass[twocolumn,showpacs,superscriptaddress,amsmath,amssymb,nofootinbib,secnumarabic,prd]{revtex4-2}
\usepackage{graphicx}
\usepackage{subfigure}
\usepackage{units}
\usepackage{multirow}
\usepackage{epsfig}
\usepackage{dcolumn}
\usepackage{booktabs}
\usepackage{appendix}
\usepackage{bm}
\usepackage{overpic}
\usepackage{color}
\usepackage{threeparttable}
\usepackage{lineno}
\usepackage{float}
\usepackage{epstopdf}
\usepackage[bookmarksnumbered, pdfstartview=FitH,colorlinks,urlcolor=blue, citecolor=blue,linkcolor=blue]{hyperref}
\usepackage[compat=1.1.0]{tikz-feynhand}

\newcommand{\BESIIIorcid}[1]{\href{https://orcid.org/#1}{\hspace*{0.1em}\raisebox{-0.45ex}{\includegraphics[width=1em]{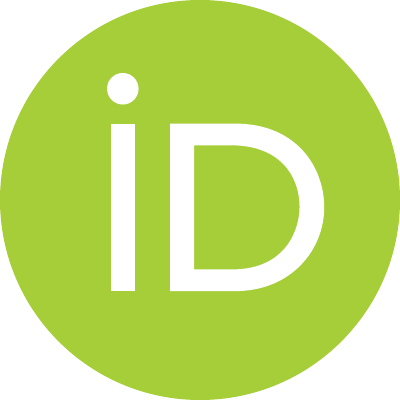}}}}

\newcommand{\gevcc}{\,\unit{GeV}/c^2}
\newcommand{\lambdacm}{\overline{\Lambda}{}_{c}^{-}}
\newcommand{\lambdacp}{\Lambda_{c}^{+}}

\newcommand{\Modea}{\bar{p}K_{S}^{0}}
\newcommand{\Modeb}{\bar{p}K^{+}\pi^-}
\newcommand{\Modec}{\bar{p}K_{S}^{0}\pi^0}
\newcommand{\Moded}{\bar{p}K_{S}^{0}\pi^-\pi^+}
\newcommand{\Modee}{\bar{p}K^{+}\pi^-\pi^0}

\newcommand{\Modeaa}{\overline{\Lambda}{}\pi^-}
\newcommand{\Modebb}{\overline{\Lambda}{}\pi^-\pi^0}
\newcommand{\Modedd}{\overline{\Lambda}{}\pi^-\pi^+\pi^-}
\newcommand{\Modeaaa}{\overline{\Sigma}{}^{0}\pi^-}
\newcommand{\Modeccc}{\overline{\Sigma}{}^{-}\pi^0}
\newcommand{\Modeddd}{\overline{\Sigma}{}^{-}\pi^-\pi^+}

\newcommand{\sgmppimev}{\Sigma^+\pi^- e^+ \nu_e}

\newcommand{\Lmdev}{\Lambda e^+ \nu_e}
\newcommand{\ppiev}{p \pi^- e^+ \nu_e}
\newcommand{\umiss}{U_{\rm miss}}

\newcommand{\results}{(2.96\pm0.95_{\rm stat}\pm0.23_{\rm syst})\times10^{-4}}

\begin{document}
\title{\bf\boldmath Evidence for the semileptonic decay \texorpdfstring{$\Lambda_c^{+} \to p \pi^{-} e^+ \nu_e$}{Lc->Sgmpiev}}
\date{\it \small \bf \today}
\author{
M.~Ablikim$^{1}$\BESIIIorcid{0000-0002-3935-619X},
M.~N.~Achasov$^{4,d}$\BESIIIorcid{0000-0002-9400-8622},
P.~Adlarson$^{81}$\BESIIIorcid{0000-0001-6280-3851},
X.~C.~Ai$^{87}$\BESIIIorcid{0000-0003-3856-2415},
C.~S.~Akondi$^{31A,31B}$\BESIIIorcid{0000-0001-6303-5217},
R.~Aliberti$^{39}$\BESIIIorcid{0000-0003-3500-4012},
A.~Amoroso$^{80A,80C}$\BESIIIorcid{0000-0002-3095-8610},
Q.~An$^{64,77,\dagger}$,
Y.~H.~An$^{87}$\BESIIIorcid{0009-0008-3419-0849},
Y.~Bai$^{62}$\BESIIIorcid{0000-0001-6593-5665}
% ...
\\
\vspace{0.2cm}
(BESIII Collaboration)\\
\vspace{0.2cm} {\it
$^{1}$ Institute of High Energy Physics, Beijing 100049, People's Republic of China\\
% ...
$^{4}$ Budker Institute of Nuclear Physics SB RAS (BINP), Novosibirsk 630090, Russia\\
% ...
$^{31}$ INFN Sezione di Ferrara, (A)INFN Sezione di Ferrara, I-44122, Ferrara, Italy; (B)University of Ferrara, I-44122, Ferrara, Italy\\
% ...
$^{39}$ Johannes Gutenberg University of Mainz, Johann-Joachim-Becher-Weg 45, D-55099 Mainz, Germany\\
% ...
$^{62}$ Southeast University, Nanjing 211100, People's Republic of China\\
% ...
$^{64}$ State Key Laboratory of Particle Detection and Electronics, Beijing 100049, Hefei 230026, People's Republic of China\\
% ...
$^{77}$ University of Science and Technology of China, Hefei 230026, People's Republic of China\\
% ...
$^{80}$ University of Turin and INFN, (A)University of Turin, I-10125, Turin, Italy; (B)University of Eastern Piedmont, I-15121, Alessandria, Italy; (C)INFN, I-10125, Turin, Italy\\
$^{81}$ Uppsala University, Box 516, SE-75120 Uppsala, Sweden\\
% ...
$^{87}$ Zhengzhou University, Zhengzhou 450001, People's Republic of China\\

\vspace{0.2cm}
$^{\dagger}$ Deceased\\
% ...
$^{d}$ Also at the Novosibirsk State University, Novosibirsk, 630090, Russia\\
% ...
}
}
\begin{abstract}
Based on $4.5\, \unit{fb^{-1}}$ of $e^+e^-$ collision data collected with the BESIII detector at the BEPCII collider at center-of-mass energies between $4.600\,\unit{GeV}$ and $4.699\,\unit{GeV}$, the first search for the Cabbibo-suppressed semileptonic decay $\lambdacp\to\ppiev$ is performed. The branching fraction of $\lambdacp\to\ppiev$ is measured to be $\results$ with a signal significance of $4.2\sigma$.
\end{abstract}

\maketitle
\section{Introduction}
Semileptonic~(SL) decays of the ground-state charmed baryon $\lambdacp$ provide a clean platform for studying the structure of the $\lambdacp$ and the final state hadrons.
These $\Lambda_c^+$ SL decays proceed via the weak interaction as either Cabibbo-favored (CF) $c\to s$ or Cabibbo-suppressed (CS) $c\to d$ transitions.
The BESIII experiment has performed several studies of both CF and CS $\lambdacp$ SL decays. Among CF SL decays, the process $\lambdacp\to\Lambda \ell^+ \nu_\ell$ ($\ell$ is $e$ or $\mu$) is dominant. BESIII has performed the most precise measurements of these branching fractions (BFs) and the first form factor measurements~\cite{BESIII:2022ysa,BESIII:2023jxv}.
Furthermore, the CF SL decays with an excited $\Lambda^{*}$, $\lambdacp\to\Lambda^{*}e^+\nu_e$, have been searched for in the $pK^-e^+\nu_e$~\cite{BESIII:2022qaf}, $\Sigma^\pm\pi^\mp e^+\nu_e$~\cite{BESIII:Sgmpienu}, $\Lambda\pi^+\pi^-e^+\nu_e$ and $pK_S^0\pi^-e^+\nu_e$~\cite{BESIII:2023jem} channels. 
Among these final states, the BF of $\lambdacp\to pK^-e^+\nu_e$ is measured to be $(0.88\pm0.17_{\mathrm{stat}}\pm0.07_{\mathrm{syst}})\times 10^{-3}$ and evidence of SL decays with intermediate $\Lambda^*$ resonances $\lambdacp\to\Lambda(1405)e^+\nu_e$ and $\lambdacp\to\Lambda(1520)e^+\nu_e$ is reported~\cite{BESIII:2022qaf}. Evidence for the decay $\lambdacp\to\Sigma^\pm\pi^\mp e^+ \nu_e$ is seen, with a statistical significance of over $3\sigma$~\cite{BESIII:Sgmpienu} while upper limits for $\lambdacp\to\Lambda\pi^+\pi^- e^+\nu_e$ and $\lambdacp\to p K_s^0 \pi^-e^+\nu_e$ decays are set~\cite{BESIII:2023jem}.
These measurements provide valuable insight into the $\Lambda_c^+\to\Lambda^{*}$ transition.
Experimentally, the study of CS SL decays is much more difficult than for CF SL decays because the former have smaller BFs and are also heavily contaminated with CF background.
The benchmark CS SL decay into a neutron $\lambdacp\to ne^+\nu_e$ has been observed with a BF of $(3.57 \pm 0.34_{\mathrm{stat.}} \pm 0.14_{\mathrm{syst.}}) \times 10^{-3}$~\cite{BESIII:2024mgg}, which motivates the search for the $\lambdacp$ SL decays with an excited nucleon $\lambdacp \to N^* e^+\nu_e$. 

Only a few theoretical studies have been conducted on the $\lambdacp \to N^* e^+\nu_e$ process. Due to the lack of experimental constraints, there is a wide range of theoretical predictions for the BF of this decay. Reference~\cite{Pervin:2005ve} uses the constituent quark model to calculate the BF of $\lambdacp \to N^*(1535) e^+\nu_e$ to be $4.03\times10^{-5}$ or $8.06\times10^{-5}$. 
Reference~\cite{Emmerich:2016jjm} predicts the BF of $\lambdacp \to N^*(1535) e^+\nu_e$ to be $6.4\times10^{-3}$ or $7.7\times10^{-4}$ using light-cone sum rules with different treatments of the $N^*$ distribution amplitude. An experimental measurement is crucial for testing different theoretical calculations and advancing the understanding of the decay dynamics of $\lambdacp\to N^{*}e^+\nu_e$.

In this paper, we search for the CS SL decay \mbox{$\lambdacp\to\ppiev$} using $e^+e^-$ collision data with a total integrated luminosity of $4.5\ \unit{fb^{-1}}$ collected at the center-of-mass energies $\sqrt{s} = 4.600$, 4.612, 4.628, 4.641, 4.661, 4.682, and 4.699~GeV~\cite{BESIII:2022ulv, BESIII:2015zbz, BESIII:2022dxl, Liao:2025lth} with the BESIII detector. Charge-conjugate channels are implied throughout this paper.
\section{BESIII detector and Monte Carlo samples}
The BESIII detector, a cylindrical spectrometer at the BEPCII electron-positron collider~\cite{Yu:2016cof}, consists of a helium-based multilayer drift chamber (MDC), a time-of-flight system (TOF), a CsI(Tl) electromagnetic calorimeter (EMC) inside a superconducting solenoidal magnet providing a 1.0 T magnetic field, and a muon counter. 
The design and performance of the BESIII detector is described in detail in 
Refs.~\cite{BESIII:2009fln_bes3, BESIII:2020nme}.

Monte Carlo (MC) simulated data samples produced with a {\sc geant4}-based~\cite{GEANT4:2002zbu,Huang:2022wuo} software package, which
includes the geometric description of the BESIII detector and the detector response, are used to determine detection efficiencies and to estimate backgrounds.
The simulation models the beam energy spread and initial state radiation (ISR) in the $e^+e^-$ annihilations with the generator {\sc kkmc}~\cite{Jadach:2000ir}. 
The inclusive MC sample is generated, simulating the $\lambdacp\lambdacm$ production and all subsequent decays along with other products of $e^+e^-$ annihilation, including open charm mesons, ISR production of vector charmonium(-like) states, and continuum processes. 
All particle decays are modeled by {\sc evtgen}~\cite{Lange:2001uf_EvtGen, Ping:2008zz_EvtGen}, with the Particle Data Group (PDG) decay BFs~\cite{ParticleDataGroup:2024cfk}. 
Final state radiation from charged final state particles is incorporated using the {\sc photos} package~\cite{Richter-Was:1992hxq_photos}. 
The signal MC sample simulates the $e^+e^-\to\lambdacp\lambdacm$, where the $\lambdacm$ decays to 11 hadronic modes, $\Modea$, $\Modeb$, $\Modec$, $\Moded$, $\Modee$, $\Modeaa$, $\Modebb$, $\Modedd$, $\Modeaaa$, $\Modeccc$, and $\Modeddd$, while the $\lambdacp$ decays in the signal mode, $\lambdacp\to p\pi^- e^+ \nu_e$, which is generated uniformly in phase space.
High-statistics MC samples for the major peaking backgrounds and a control sample are generated for background estimation and validation, including $\lambdacp\to\Lmdev$, $\lambdacp\to\sgmppimev$, $\lambdacp\to\Sigma^+\pi^-\pi^+$ and $\lambdacp\to p \pi^- \pi^+$. 
\section{Analysis}
In data samples collected near the $\lambdacp\lambdacm$ pair production threshold, $\lambdacp\lambdacm$ pairs are produced without other hadronic companions. Reconstructing a single-tag (ST) $\lambdacm$ with a known hadronic decay mode, combined with the known beam energy, provides the full four-momentum of the partially reconstructed SL $\lambdacp$ decay. This technique, first introduced by the MARK-III collaboration~\cite{MARK-III:1985hbd}, is called the double tag (DT) method, . 

The ST decay modes are reconstructed with the same methods and requirements as in the previous BESIII analysis~\cite{BESIII:Sgmpienu}. The ST yields are obtained by fitting the distributions of the beam-constrained mass
\begin{linenomath*}
    \begin{eqnarray}
        M_{\rm BC} = \sqrt{E_{\rm beam}^2/c^4 - |\vec{p}|^2/c^2},
    \end{eqnarray}
\end{linenomath*}
where $E_{\rm beam}$ is the beam energy and $\vec{p}$ is the ST $\lambdacm$ momentum in the initial $e^+e^-$ rest frame. The ST yields are counted in the $M_{\rm BC}$ signal region, (2.28, 2.30)$\gevcc$. In total, $116154\pm444$ ST candidates are found among the $\lambdacp\lambdacm$ pairs. 

In the signal decay $\lambdacp \to p \pi^- e^+ \nu_e$, the neutrino can not be directly detected in the BESIII detector. To identify the SL signals, a kinetic variable, $\umiss$, is defined as
\begin{linenomath*}
    \begin{eqnarray}
        U_{\mathrm{miss}} = E_{\mathrm{miss}} - c|{\vec{p}_{\mathrm{miss}}}|,
    \end{eqnarray}
\end{linenomath*}
where $E_{\mathrm{miss}}$ is the missing energy calculated with $E_{\mathrm{miss}} = E_{\mathrm{beam}} - E_{p} - E_{\pi^-} - E_{e^+}$ and the missing momentum $\vec{p}_{\mathrm{miss}}$ is calculated with $\vec{p}_{\mathrm{miss}} = \vec{p}_{\lambdacp} - \vec{p}_{p} - \vec{p}_{\pi^-} - \vec{p}_{e^+}$. The momentum of the $\lambdacp$ is inferred by the kinematic information of the exclusively reconstructed ST $\lambdacm$, $\vec{p}_{\lambdacp}=-\hat{p}_{\mathrm{tag}}\sqrt{E^{2}_{\mathrm{beam}}/c^{2}-m^{2}_{\lambdacm}c^{2}}$, where $\hat{p}_{\mathrm{tag}}$ is the  ST $\lambdacm$ momentum direction and $m_{\lambdacm}$ is the known mass of the $\lambdacm$~\cite{ParticleDataGroup:2024cfk}. The $\umiss$ values peak at zero for signal events. The BF of signal decay can be calculated with
\begin{linenomath*}
    \begin{eqnarray}
        \mathcal{B} = \frac{N^{\rm DT}}{N^{\rm ST} \, \varepsilon},
        \label{eq:bf}
    \end{eqnarray}
\end{linenomath*}
where $N^{\mathrm{DT}}$ is the DT yield of the signal decay, $N^{\mathrm{ST}}$ is the total yield of ST $\lambdacm$ signals and $\varepsilon$ is the detection efficiency. The detection efficiency is averaged across all ST modes and energy points, calculated by 
\begin{linenomath*}
    \begin{eqnarray}
        \varepsilon = \sum_{ij}\left(\frac{N_{ij}^{\rm ST}}{\varepsilon_{ij}^{\rm ST}} \, \varepsilon_{ij}^{\rm DT}\right)/N^{\rm ST},
        \label{eq:bf2}
    \end{eqnarray}
\end{linenomath*}
where $i$ and $j$ represent different ST modes and energy points, respectively.

The signal decay $\lambdacp \to p \pi^- e^+ \nu_e$ is reconstructed from the remaining particles recoiling against the ST $\lambdacm$. We require that there are exactly three good charged tracks which satisfy $|\cos\theta|<0.93$, and have a distance of closest approach to the interaction point (IP) less than 10 cm along the $z$-axis and less than 1 cm in the transverse plane. Here, $\theta$ is the angle between the track and the $z$-axis (the symmetry axis of the MDC). 
Particle identification (PID) is performed by calculating the likelihood of a charged track being a proton, kaon, or pion with the energy deposited in the MDC ($dE/dx$) and the time of flight in the TOF. For identification of positrons, information from the EMC is added.
We identify the pion, proton and positron candidates by comparing the likelihoods of the hypotheses. 
If $\mathcal{L}(\pi)>0$, $\mathcal{L}(\pi) > \mathcal{L}(K)$, tracks are identified as pions. If $\mathcal{L}(p)>0$, $\mathcal{L}(p) > \mathcal{L}(\pi)$ and $\mathcal{L}(p) > \mathcal{L}(K)$, tracks are identified as protons. If $\mathcal{L}(e)>0$ and $\mathcal{L}(e)/(\mathcal{L}(e) + \mathcal{L}(\pi) + \mathcal{L}(K))>0.8$, tracks are identified as positrons. 
A vertex fit is performed assuming the pion, proton and positron candidate tracks come from the same vertex. The $\chi^2$ of the vertex fit is required to satisfy $\chi^2_{\rm vertex}<22$.

The $\lambdacp\to\Lambda e^+ \nu_e, \Lambda\to p \pi^-$ decay contains the same final state particle as the signal decay with a much larger BF, which makes it a major background source. A vertex fit, for the proton and pion, and a secondary vertex fit, for that system to originate from the IP, are performed with assumption that the proton and pion candidates come from a $\Lambda$ decay. The decay length significance $L$ and its uncertainty $\sigma_L$ are calculated. To reduce the $\lambdacp\to\Lambda e^+ \nu_e$ background, the vertex-constrained $p\pi^-$ invariant mass $M_{p\pi^-}$ is required to be larger than $1.13\gevcc$ and the decay length is required to satisfy $|L/\sigma_L|<3$. Another background is introduced where a pion is misidentified as the positron candidate. The decays $\lambdacp \to p \pi^+ \pi^-$ and $\lambdacp\to p K_S^0, K_S^0 \to \pi^+ \pi^-$ with similar final state as the signal decay contribute dominantly to this background. We require $M_{p\pi^-\pi(e)^+}<2.27\gevcc$ to reject these backgrounds, where $M_{p\pi^-\pi(e)^+}$ is the invariant mass of the $p\pi^- e^+$ candidate calculated by replacing the positron mass with the pion mass. Most of the remaining backgrounds contain a $\pi^0$ in the final state. We require the opening angle between the positron and pion candidate $\cos\theta_{\pi,e}<0.99$ to suppress background from $\gamma$-conversions. The above requirements are optimized with the Punzi figure-of-merit (FOM)~\cite{Punzi:2003bu}, $\frac{\varepsilon}{5/2+\sqrt{B}}$, where $\varepsilon$ is the detection efficiency and $B$ is the number of background events in the inclusive MC sample.

A boosted decision tree (BDT) classifier is trained to suppress the remaining backgrounds, most of which contain an extra $\pi^0$. Five input variables are used in the training: the likelihood ratio of the electron PID, the total energy of the good showers, the cosine of the angle between the missing momentum and the momentum of the largest-energy shower, the number of $\pi^0$ candidates that satisfy $M_{\gamma\gamma}\in(0.098,1.165)\gevcc$, and the $M_{\rm BC}$ of the ST side $\lambdacm$.
These variables are chosen because they are not sensitive to the $\lambdacp\to\ppiev$ signal models, while still capturing the signature difference between signal and background.
The signal MC sample is used as the signal training sample while the background training sample is the inclusive MC sample without the $\lambdacp\to\Lambda e^+ \nu_e$ component. The consistency between MC sample and data is validated with the $\lambdacp\to\Lmdev$ control sample. The requirement on the BDT output is obtained via optimization with a Punzi FOM.

\begin{figure}[tpb]
    \centering
    \includegraphics[width=0.9\linewidth]{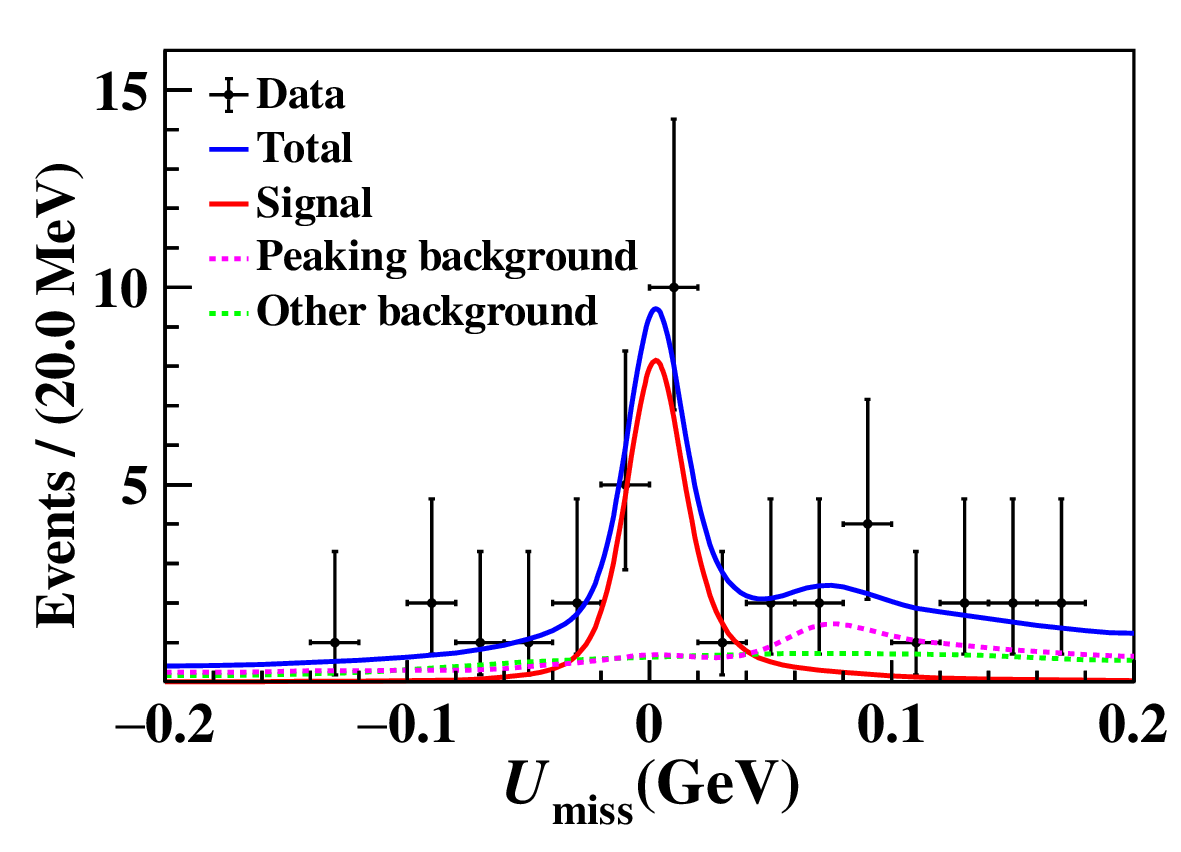}
    \caption{The $U_{\rm miss}$ distribution of the $\lambdacp\to\ppiev$ signal candidates in data and the corresponding fit results. The solid blue line is total fit curve. The solid red line describes signal shape. The dashed magenta line indicates peaking background, which include the $\lambdacp\to\Sigma^+\pi^+\pi^-$, $\lambdacp\to\Lmdev$ and $\lambdacp\to\sgmppimev$ processes. The dashed green line represents other background.}
    \label{fig:data}
\end{figure}

The average DT efficiency after all requirements is 43.7\%. The DT yield is extracted with an unbinned maximum likelihood fit. The signal shape is modeled by the MC-simulated shape convolved with a Gaussian function representing the difference between data and MC sample. The parameters of the Gaussian function are extracted from the $\lambdacp \to \Lmdev$ control sample. The remaining peaking backgrounds $\lambdacp\to\Sigma^+\pi^+\pi^-$, $\lambdacp\to\Lmdev$ and $\lambdacp\to\sgmppimev$ are described with MC simulated shapes. Their yields are fixed to the estimated values based on the known BFs~\cite{ParticleDataGroup:2024cfk}. The other backgrounds are divided into two categories: $\lambdacp$ and non-$\lambdacp$ decays. Their shapes are described by MC simulation and their magnitudes are free parameters in the fit. The fit results are shown in Fig.~\ref{fig:data}. The DT yield is measured to be $15.4\pm4.9$, which gives a BF of $(2.96\pm0.95)\times 10^{-4}$ with a $4.4\sigma$ statistical significance. The statistical significance is estimated using $\sqrt{-2 \, \Delta\mathcal{L}}$, where $\Delta\mathcal{L}$ is the likelihood difference of the fit with and without including the signal component. 

\section{Systematic Uncertainty}
The systematic uncertainties from ST $\bar{\Lambda}_c^-$ selections are largely canceled in the DT method. The remaining systematic uncertainties mainly come from tracking, PID, and other requirements applied in selecting the signal $\Lambda_c^+$.  The sources of the systematic uncertainties are listed in Table~\ref{tab:sys_err}.

\begin{table}[tpb]
  \begin{center}
     \caption{Systematic uncertainties in the BF measurement.}
  \begin{tabular}{l|c}
    \hline \hline
    	Source           & Uncertainty (\%) \\  \hline
	BDT              & 2.6 \\
	$\cos\theta_{\pi,e}<0.99$ & negligible \\
	$\Lmdev$ veto    & 3.9 \\
	$M_{p\pi^-\pi(e)^+}<2.27\gevcc$ & 0.1 \\	
	$e^+$ tracking   & 0.5 \\ 
        $e^+$ PID        & 1.9 \\
	Proton tracking  & 0.6 \\
        Proton PID       & 0.7 \\
	$\pi^-$ tracking & 0.2\\
	$\pi^-$ PID      & 0.2\\  
        MC signal model  & 4.8 \\ 
	MC statistics    & 0.3 \\
	ST $\lambdacm$ yield & 1.0 \\
        Fit &  2.8\\
    \hline
	Total  & 7.7 \\
    \hline \hline
   \end{tabular}
   \label{tab:sys_err}
  \end{center}
  \end{table}

The systematic uncertainties of the BDT requirement and the $\cos\theta_{\pi,e}<0.99$ requirement are studied with the $\lambdacp\to\Lmdev$ control sample. The event selection applied to control sample is similar to that used in Ref.~\cite{BESIII:2022ysa}, with the same ST modes as in this analysis. The MC sample for $\lambdacp\to\Lmdev$ is modeled with the measured form factor results in Ref.~\cite{BESIII:2022ysa}. We measure the BF of $\lambdacp\to\Lmdev$ with and without applying the BDT or $\cos\theta_{\pi,e}<0.99$ requirement. The relative difference between the measured BFs with and without the requirement is taken as the systematic uncertainty. The measured BF with the BDT requirement applied is consistent with the PDG value, which further validates the method.

The $\lambdacp\to p \pi^+\pi^-$ control sample is used to evaluate the other requirements applied to veto the $\lambdacp\to\Lmdev$ background. The efficiency of the $\Lmdev$ veto is estimated by the combined efficiency of $M_{p\pi^-}>1.13\gevcc$, $|L/\sigma|<3$ and $\chi^2_{\rm vertex}<22$ requirements, evaluated using both data and MC samples. The relative difference in efficiency is taken as the systematic uncertainty. For the $M_{p\pi^-\pi(e)^+}<2.27\gevcc$ requirement, the resolution difference of the $M_{p\pi^-\pi^+}$ distribution between data and MC sample is extracted from the control sample as a Gaussian function. Then the $M_{p\pi^-\pi(e)^+}$ distribution of the $\lambdacp\to\ppiev$ signal MC sample is smeared with this Gaussian function. The systematic uncertainty is estimated by the efficiency difference before and after the Gaussian smearing.

The systematic uncertainties of tracking and PID for electrons, protons and pions are studied with \mbox{$e^+e^-\to e^+e^-\gamma$}, \mbox{$J/\psi\to p \bar{p} \pi^+\pi^-$} and \mbox{$e^+e^-\to K^+K^-\pi^+\pi^-$}, $K^+K^-\pi^+\pi^-\pi^0$, $\pi^+\pi^-\pi^+\pi^-$, $\pi^+\pi^-\pi^+\pi^-\pi^0$ control samples, respectively. The efficiency differences are obtained in different momentum regions. The systematic uncertainties are estimated by re-weighting the signal MC sample with the these efficiency differences. The MC model is changed from a uniform phase-space model to a model with an $N^*$ resonance to evaluate the systematic uncertainties. The lowest nucleon excited state, $N^*(1440)$, and one with a theoretical prediction, $N^*(1535)$, are considered. The largest efficiency difference of MC models with $N^*$ states is taken as the systematic uncertainty due to the MC model. 

The systematic uncertainty arising from the limited statistics of the MC samples is estimated to be $0.3\%$. The systematic uncertainty of the ST $\lambdacm$ yields is assigned to be $1\%$~\cite{BESIII:2022ysa}.

The systematic uncertainty of the fit is estimated by changing the descriptions of different components in the fit model. The MC-simulated signal shape is changed to a double-side Crystal Ball function. The fixed peaking background yields are varied by one standard deviation. The shape for the hadron background is modeled by the $M_{\rm BC}$ sideband of ST $\lambdacm$ instead of the MC shape. The sum in quadrature of the various contributions is taken as the total systematic uncertainty. 
These systematic uncertainties are the only additive sources, which affect the evaluation of statistical significance.
The lowest statistical significance given by the different fit models used is taken as the statistical significance considering the systematic uncertainty, which is $4.2\sigma$.
\section{Summary}

In summary, using $4.5\,\unit{fb^{-1}}$ of $e^+e^-$ collision data collected at center-of-mass energies from 4.600 to 4.699 GeV with the BESIII detector, we present a study of the SL decay $\lambdacp \to \ppiev$. 
Evidence for $\lambdacp \to \ppiev$ is found with a signal significance of $4.2\sigma$. The BF is measured to be $\mathcal{B}(\lambdacp \to \ppiev) = \results$. 
\begin{figure}[tpb]
    \centering
    \includegraphics[width=0.9\linewidth]{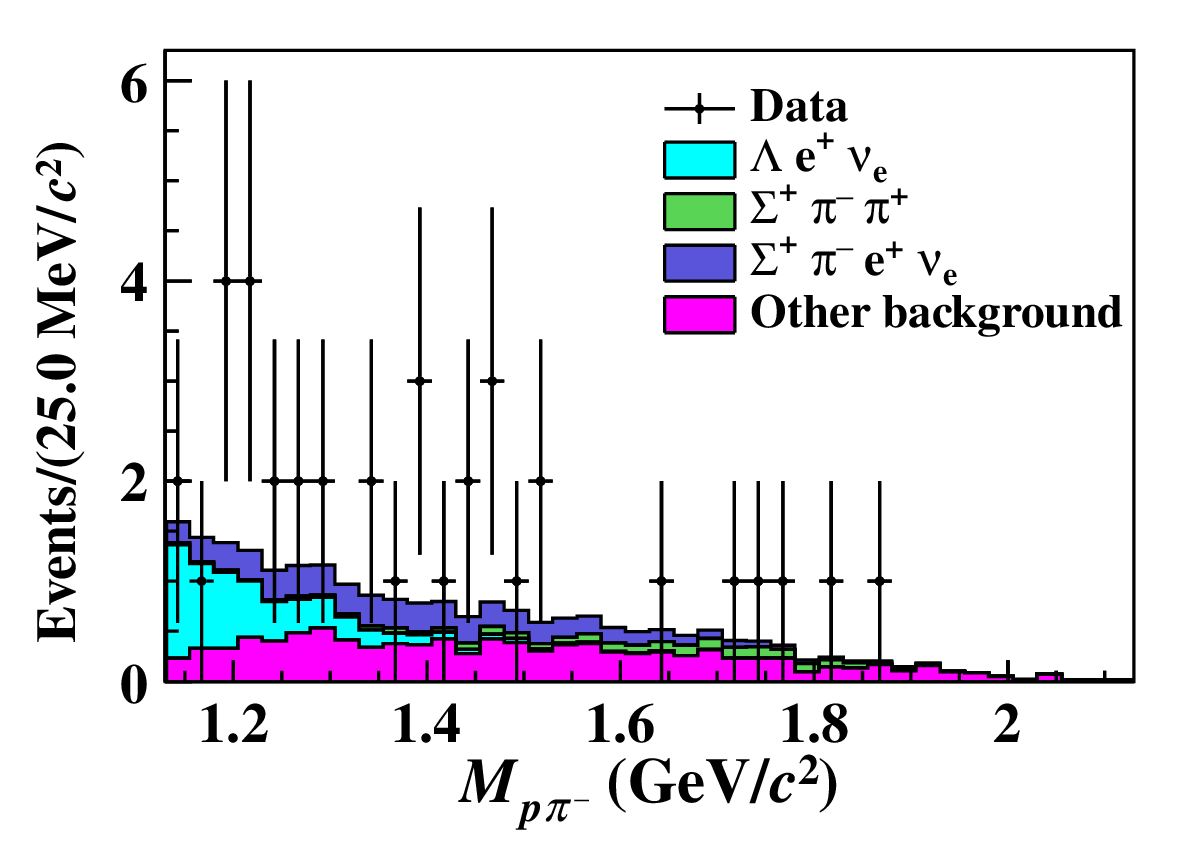}
    \caption{The $M_{p\pi^-}$ distribution of the $\lambdacp\to\ppiev$ signal candidates in data. The histograms show the background distribution estimated with MC shapes and yields from signal fit results.}
    \label{fig:Mppi}
\end{figure}
The $M_{p\pi^-}$ distribution of the data candidates are shown in Fig.~\ref{fig:Mppi}, with background estimated by MC shapes and yields from the signal fit results. 
There is no evidence of intermediate states in the $M_{p\pi^-}$ spectrum with the current statistics.   
Our study provides a constraint on the theoretical calculations of the BFs and form factors of $\Lambda_c^+\to N^{*}e^+\nu_e$. 
The larger prediction of the $\Lambda_c^+\to N^{*}(1535)e^+\nu_e$ BF by Ref.~\cite{Emmerich:2016jjm} is disfavored by our results, since the prediction is too large, even if all $p\pi^-$ is assumed to be from the $N^*(1535)$. Larger data samples collected at BESIII in the near future~\cite{BESIII:2020nme,Li:2021iwf} will significantly enhance the sensitivity of this analysis, enabling the isolation of contributions from intermediate excited nucleon states and paving the way for a deeper understanding of charmed baryon decays.

\textbf{Acknowledgement}

The BESIII Collaboration thanks the staff of BEPCII (https://cstr.cn/31109.02.BEPC) and the IHEP computing center for their strong support. This work is supported in part by National Key R\&D Program of China under Contracts Nos. 2023YFA1606000, 2023YFA1606704; National Natural Science Foundation of China (NSFC) under Contracts Nos. 11635010, 11935015, 11935016, 11935018, 12025502, 12035009, 12035013, 12061131003, 12192260, 12192261, 12192262, 12192263, 12192264, 12192265, 12221005, 12225509, 12235017, 12342502, 12361141819; the Chinese Academy of Sciences (CAS) Large-Scale Scientific Facility Program; the Strategic Priority Research Program of Chinese Academy of Sciences under Contract No. XDA0480600; CAS under Contract No. YSBR-101; 100 Talents Program of CAS; The Institute of Nuclear and Particle Physics (INPAC) and Shanghai Key Laboratory for Particle Physics and Cosmology; ERC under Contract No. 758462; German Research Foundation DFG under Contract No. FOR5327; Istituto Nazionale di Fisica Nucleare, Italy; Knut and Alice Wallenberg Foundation under Contracts Nos. 2021.0174, 2021.0299, 2023.0315; Ministry of Development of Turkey under Contract No. DPT2006K-120470; National Research Foundation of Korea under Contract No. NRF-2022R1A2C1092335; National Science and Technology fund of Mongolia; Polish National Science Centre under Contract No. 2024/53/B/ST2/00975; STFC (United Kingdom); Swedish Research Council under Contract No. 2019.04595; U. S. Department of Energy under Contract No. DE-FG02-05ER41374


\bibliographystyle{apsrev4-2}
\bibliography{ref}
\end{document}